\documentclass[11pt,a4paper]{article}
\usepackage[T1]{fontenc}
\usepackage[utf8]{inputenc}
\usepackage[english]{babel}
\usepackage{lmodern}
\usepackage{microtype}
\usepackage{geometry}
\usepackage{setspace}
\usepackage{csquotes}
\usepackage{hyperref}
\hypersetup{colorlinks=true, urlcolor=blue, linkcolor=black, citecolor=black}
\usepackage{enumitem}
\setlist{noitemsep}
\usepackage{amsmath,amssymb}
\usepackage{pifont}
\usepackage{footmisc}

\title{Lorentz, Poincaré, Einstein, and the Genesis of the Theory of Special Relativity}
\author{Hector Giacomini\\
Institut Denis Poisson. Universit\'e d'Orl\'eans -- Universit\'e de Tours -- CNRS.\\
37200 Tours, France.\\
\texttt{giacominihector@gmail.com}}
\date{}
\begin{document}
\maketitle
\onehalfspacing

\begin{abstract}
This article reexamines the genesis of special relativity by situating the contributions of Lorentz, Poincaré, and Einstein within the scientific, documentary, and editorial context of the years 1895--1913.  
It emphasizes the rapid circulation of Lorentz’s 1904 work, in particular through German-speaking channels such as the \textit{Beiblätter zu den Annalen der Physik}, and reassesses the significance of Richard Gans’s 1905 review as a concise access point to Lorentz’s results.  
The article also discusses Poincaré’s role in formulating the principle of relativity, interpreting local time, establishing the group property of the Lorentz transformations, and developing an invariant formulation of electrodynamics.  
Against this background, Einstein’s 1905 paper appears not as an isolated creation, but as a powerful reformulation of problems already posed by Maxwellian electrodynamics and by the failure to detect motion through the ether.  
The article finally examines the subsequent construction of the Lorentz--Einstein--Minkowski canon and the relative exclusion of Poincaré from that narrative.  
Its central claim is that special relativity should be understood as the crystallization of a broader electrodynamic transformation of physics, rather than as a sudden break detached from its immediate scientific context.
\end{abstract}

\section{The University of Bern and its Library in 1905}

Since the theory of special relativity is associated with Albert Einstein, who settled in Bern in 1902 and remained there until 1909, it is useful to recall the history of the University and its library.

The University of Bern, officially founded in 1834 \cite{unibe-hist}, belongs to a long tradition of higher education in the city dating back to the sixteenth century.  
In 1528, Bern established a High School intended for the training of Protestant pastors.  
Transformed into an Academy in 1805, it comprised four faculties: theology, law, medicine, and philosophy.  
Under the impulse of the liberal Bernese government, the Academy became a university in 1834, conceived in a spirit of social and political openness.  
The institution experienced rapid growth, especially after the creation of the Swiss federal state in 1848.  
By the late nineteenth century, it had become one of the leading universities in Switzerland.  
Foreign students, particularly from Germany and Russia, made up a large part of the enrollment.  
Among them, Russian women students played an important role in the admission of women to university studies in Bern.  
At the turn of the century, Bern also distinguished itself through several remarkable figures: in 1903, the philosopher Anna Tumarkin became the first woman in Europe authorized to supervise doctoral theses and to serve on a university council.  
In the same year, the university inaugurated its main building on the Grosse Schanze, designed by the Bernese architects Alfred Hodler and Eduard Joos.  
A few years later, in 1909, the surgeon Theodor Kocher, a professor at the institution, was awarded the Nobel Prize in Medicine, further strengthening the university’s international reputation.

The University Library \cite{ub-geschichte} had deep roots in the city’s scholarly institutions.  
After the Reformation, the collections of the Latin school, the collegiate church, and the dissolved Bernese monasteries were brought together between 1533 and 1535.  
In the nineteenth century, the library system underwent several reorganizations, and by the beginning of the twentieth century it had become one of the principal documentary resources of the city.  
The library was not a closed university facility.  
It was open to students, professors, members of learned societies, and, under regulated conditions, to a wider public of readers.  
This openness is important for understanding the documentary environment available in Bern around 1905.

The services included direct lending, on-site consultation in the reading rooms, and access to materials not held locally through interlibrary exchange.  
Readers in Bern could therefore consult not only the library’s own holdings but also, when necessary, request books or journals from other Swiss university libraries.  
This system provided access to a scientific collection broader than that physically present in Bern itself.

Opening hours were generous \cite{ub-bern-oeffnungszeiten}.  
The lending service was open from 10 a.m. to 12 p.m. and from 2 p.m. to 4 p.m., but the reading rooms remained open longer: from 10 a.m. to 12 p.m. and from 2 p.m. to 7 p.m., and until 8 p.m. in winter.  
The reading rooms contained around 300 journals and numerous reference works.  
This distinction is important: even after the lending desk had closed, a reader could still consult periodicals and reference material on site. 
The library was closed on Sundays and public holidays, as well as during Easter week and in September; outside these periods, even during the academic vacations, it remained accessible for a few hours a day according to notices posted at the entrance.

In 1904--1905, a new building was being prepared on M\"unstergasse to accommodate the growing collections.  
The reorganization and modernization of the library infrastructure marked an important step for the scholarly life of Bern.  
In this setting, the University Library was not a marginal institution but a central resource for anyone in the city wishing to follow current scientific literature.

\section{The Role of the Journal \textit{Beibl\"atter zu den Annalen der Physik}}

In 1905, the \textit{Beibl\"atter zu den Annalen der Physik} occupied an essential position within the German-speaking scientific system.  
Conceived as a bibliographical supplement to the \textit{Annalen der Physik}, the journal did not publish original research but compiled analytical summaries, critical reviews, and systematic listings of international literature in physics, chemistry, and related disciplines.  
Its primary mission was to provide an information service offering a continuous overview of scientific production worldwide.

Edited in Leipzig, the \textit{Beibl\"atter} were directed by Eilhard Wiedemann until 1900, and subsequently by Walter K\"onig between 1901 and 1907.  
A cumulative index covering volumes 24 to 30 (1900--1906) still allows researchers to identify reviewers and cited authors, and remains a valuable research tool for reconstructing reading and dissemination networks.

Within the German-speaking world, the \textit{Beibl\"atter} fulfilled a very important function.  
For German-speaking physicists, they served as a rapid information tool enabling them to follow an expanding international literature.  
Students and researchers could use them as an entry point to specialized journals, which were often difficult to access directly.  
University libraries also relied on such bibliographical notices to orient acquisitions and requests, particularly for costly foreign journals.  
Their monthly publication ensured a regular and up-to-date flow of information.

In Switzerland, the \textit{Beibl\"atter} were available in the main university libraries, including Zurich, Geneva, Basel, and Bern.  
For institutions with more modest means than the largest German universities, they represented a major resource: instead of multiplying expensive subscriptions to foreign journals, these libraries could rely on this centralized repertory to identify relevant articles and, when necessary, request them through interlibrary exchange.  
Their transnational diffusion extended beyond the German-speaking area, since the \textit{Beibl\"atter} formed part of a wider European system of scientific information.

In 1905, the \textit{Beibl\"atter zu den Annalen der Physik} were part of the collections of the University Library of Bern, where they were accessible to readers interested in current scientific literature.  
This is a significant point for the present study: a reader in Bern concerned with electrodynamics could consult not only original physics journals such as the \textit{Annalen der Physik}, but also a review periodical that summarized, classified, and indexed international work in the field.  

\section{Maxwell’s Equations: A Scientific and Technological Revolution}

\subsection{The Unification of Electricity, Magnetism, and Light}

Between 1861 and 1873, James Clerk Maxwell accomplished an unprecedented intellectual synthesis.  
This synthesis, however, was part of a longer historical lineage.  
André-Marie Ampère had already formulated in the 1820s the quantitative laws of the forces between electric currents, laying the foundations of electrodynamics.  
From 1831 onward, Michael Faraday discovered electromagnetic induction and introduced the notion of lines of force, which directly inspired Maxwell.  
Maxwell himself acknowledged that his theory was, to a large extent, a mathematical translation and generalization of Faraday’s ideas.

By formulating a coherent system of field equations, Maxwell showed that electricity and magnetism are two aspects of a single physical reality.  
More than that, he revealed that light itself is an electromagnetic wave propagating in vacuum at a speed determined by two electromagnetic constants, the electric permittivity $\varepsilon_{0}$ and the magnetic permeability $\mu_{0}$.

Maxwell thus introduced a new and decisive concept: that of the electromagnetic field.  
The field became an autonomous physical entity, capable of storing and transporting energy and momentum through space.  
This was a profound revolution.  
Physics could no longer be understood simply as a science of instantaneous action at a distance; it became increasingly a science of fields evolving in space and time.  
For a detailed historical analysis of this development, see \cite{Darrigol2000}.

\subsection{The Emergence of a Universal Velocity}

From Maxwell’s equations, in empty space, it follows that the electric and magnetic fields satisfy wave equations.  
These equations contain a propagation speed given by:
\[
c = \frac{1}{\sqrt{\mu_{0} \varepsilon_{0}}} .
\]
From the experimentally determined values of these two constants, one finds a speed equal to that of light.  
For the first time, a fundamental physical theory contained within its own structure a distinguished velocity, determined not by the motion of a material source but by electromagnetic constants.

This result created a deep tension with classical mechanics.  
Galilean kinematics was based on the ordinary addition of velocities.  
If light was an electromagnetic wave governed by Maxwell’s equations, then its speed appeared fixed by the equations themselves.  
But fixed with respect to what frame?  
In the nineteenth-century interpretation, this question was usually answered by invoking the ether as the privileged medium of propagation.  
The difficulty was that optical and electromagnetic experiments increasingly failed to reveal any motion of the Earth with respect to such a medium.

It was this deep contradiction between Maxwellian electrodynamics, the ether framework, and Galilean kinematics that opened the way to special relativity.  
Lorentz, Poincaré, and Einstein would each confront, in different ways, the problem raised by Maxwell’s theory: how could a theory containing a distinguished velocity be reconciled with the relativity of inertial motion?

Heinrich Hertz provided the most spectacular experimental confirmation of Maxwell’s prediction: he generated and detected electromagnetic waves in the laboratory.  
In less than twenty-five years, Maxwell’s theory had become one of the central pillars of physics.  
Its consequences were not only theoretical but also technological.  
From the end of the nineteenth century onward, applications derived from electromagnetism transformed communication, electrical engineering, measurement, and later much of modern technology.  
In this sense, Maxwell’s electrodynamics was not merely a brilliant theory; it reshaped both physics and technological civilization.

\subsection{Maxwell as a Model for the Great Theories of the Twentieth Century}

Maxwell’s electrodynamics also provided a powerful model for later field theories.  
It showed that a physical interaction could be described by a field with its own equations, its own energy, and its own mode of propagation.  
This conception deeply influenced twentieth-century physics.

\begin{itemize}
  \item \textbf{General Relativity}: gravitation is described not as an instantaneous force but through the geometry of space-time, in a field-theoretic framework that partly inherits the conceptual transformation initiated by Maxwell.
  \item \textbf{Quantum Field Theory}: the quantization of the electromagnetic field gave rise to quantum electrodynamics, one of the most precise theories in the history of physics.
  \item \textbf{Particle Physics}: Maxwell’s electrodynamics became one of the historical roots of gauge-theoretic thinking, later generalized in the electroweak and strong interactions.
\end{itemize}

Its impact was also conceptual and methodological.  
The field became an autonomous physical reality; interactions were no longer conceived primarily as instantaneous actions at a distance but as processes propagating through space and time.  
This shift changed the language, methods, and ambitions of theoretical physics.

Electromagnetism is therefore not merely a brilliant episode in nineteenth-century physics.  
It is one of the structural pillars of modern science and one of the main theoretical sources from which the crisis leading to special relativity emerged.

\section{The Michelson--Morley Experiments}

At the end of the nineteenth century, physics seemed securely founded on two great pillars: Newtonian mechanics and Maxwell's electrodynamics.  
Each theory had proved its validity within its own domain: mechanics for describing the motion of celestial and terrestrial bodies, electrodynamics for explaining the propagation of light waves and electrical phenomena.  
Yet their combination produced a serious difficulty.  
When Maxwellian light waves were interpreted within a mechanical ether framework, the motion of the Earth through the ether seemed to imply observable optical effects.  
The most famous attempt to detect such effects was the Michelson--Morley experiment \cite{michelson-morley1887}.

The question concerned the propagation of light.  
According to electrodynamics, light is an electromagnetic wave propagating at the speed $c$.  
In the nineteenth-century interpretation, waves were generally assumed to require a medium; for electromagnetic waves, physicists therefore postulated the existence of the ether, filling all space.  
As the Earth moves in its orbit around the Sun at about 30~km/s, it should have been moving through this medium, and that motion was expected to produce measurable effects.

The classical calculation was straightforward in spirit.  
If the laboratory moved through a stationary ether, the time taken by light to travel along an interferometer arm parallel to the Earth's motion should differ from the time taken along a perpendicular arm.  
When the apparatus was rotated, this difference should change, producing a displacement of the interference fringes.  
For the 1887 experiment, the expected shift was of the order of several tenths of a fringe, large enough to be detectable with the apparatus.

Albert A. Michelson, a physicist of Polish origin who became an American citizen, undertook a first experimental test of the ether-drift hypothesis in 1881.  
Michelson would later receive the Nobel Prize in 1907, becoming the first American laureate in physics.  
In collaboration with Edward W. Morley, an American chemist and physicist, he improved the apparatus and carried out the celebrated 1887 experiment.  
The interferometer was based on a simple principle: splitting a light beam in two, sending the parts along perpendicular arms, and then recombining them to look for possible interference shifts.  
If the Earth were moving through the ether, the light travel times along the two arms should vary with orientation, and a periodic shift in the interference fringes should appear as the apparatus was rotated.

The observed result was strongly negative.  
The fringes were not perfectly stable, and the measurements contained experimental noise, but the systematic effect predicted by the ether-drift calculation was absent.  
The observed displacements were far smaller than the expected fringe shift and could not be interpreted as the direct optical signature of the Earth's orbital motion through a stationary ether.  
Michelson and Morley therefore produced not a simple measurement of the ether wind, but a serious and persistent anomaly for the mechanical interpretation of Maxwellian optics.

This null result quickly became a major topic of discussion.  
In the following decades, commentaries, analyses, and new experimental attempts appeared in European and American scientific journals.  
The experiment could not easily be dismissed as an accidental failure: it was carefully designed, and its negative result fitted into a broader pattern of unsuccessful attempts to detect the Earth's motion through the ether.  
The mechanical framework suggested that the composition of velocities should apply to light waves as to other motions, so that optical propagation in the laboratory should depend on the Earth's motion through the privileged medium.  
The electrodynamic framework, on the other hand, fixed the speed of light through Maxwell's equations, usually understood as equations written in the ether frame.  
Combining the two frameworks made an interference shift seem natural.  
Yet the expected shift was not observed.

Physicists therefore faced a serious dilemma.  
Either Newtonian kinematics was not universally valid, and the ordinary composition of velocities did not apply in the expected way to optical phenomena; or Maxwell's electrodynamics, or its interpretation in terms of a stationary ether, had to be modified.  
The difficulty was that both mechanics and electrodynamics had been confirmed with extraordinary success within their respective domains.  
Mechanics explained planetary and terrestrial motion with remarkable accuracy; electrodynamics accounted for optical and electrical phenomena with great power.  
Abandoning either framework seemed highly problematic.

The negative result of Michelson and Morley thus became one of the major theoretical problems at the turn of the century.  
Their publications, and the reactions they provoked, show how a null result could exert a profound pressure on theory.  
Reproduced, discussed, and reinterpreted in major journals and textbooks, the experiment became known in Germany, France, Britain, and the United States.  
It did not by itself force a unique conclusion, but it made increasingly difficult any simple picture of a detectable ether wind.

Thus, the Michelson--Morley experiment became one of the symbols of the conceptual crisis of fin-de-siècle physics.  
What remained unexplained was not merely a missing fringe shift, but the apparent impossibility of detecting the Earth's motion through the supposed optical medium.  
Reconciling this fact with Maxwellian electrodynamics and the relativity of inertial motion would become one of the central challenges leading to special relativity.

\section{The Trouton--Noble Experiment}

In 1903, the British physicists Frederick Trouton and Henry R. Noble carried out an experiment designed to detect a mechanical torque acting on a charged capacitor moving through the ether \cite{trouton-noble1903}.  
Their reasoning was as follows.  
At rest, the energy of a charged capacitor is purely electrostatic and independent of its orientation.  
But if one assumes the existence of a privileged ether frame, the motion of the capacitor through this medium should modify the electromagnetic state of the system and produce a torque tending to orient the capacitor in a definite direction relative to the ether wind.

To test this prediction, Trouton and Noble suspended a charged capacitor from an extremely fine wire, in the manner of a torsion-balance experiment.  
The sensitivity of the apparatus was intended to detect very small mechanical effects.  
Despite this precision, no detectable torsion was observed.

This result added an important independent constraint to the series of ether-drift experiments.  
Unlike Michelson--Morley, which was optical and interferometric, Trouton--Noble involved an electrostatic system and a possible mechanical torque.  
Its negative result therefore broadened the problem: motion relative to the ether seemed to produce no measurable systematic optical or mechanical effect.  
Together with Michelson--Morley and related experiments, it made the hypothesis of a detectable privileged medium increasingly difficult to maintain.

\section{Hendrik Antoon Lorentz’s Contributions to the Theory of Special Relativity}

\subsection{Lorentz’s 1895 Theory}

In his 1895 memoir \cite{lorentz1895}, the Dutch physicist Hendrik Antoon Lorentz, who would later receive the Nobel Prize in Physics in 1902, sought to understand why optical phenomena seemed independent of the Earth’s motion through the ether.  
He proposed an approximate transformation of space and time coordinates, valid to first order in \(v/c\), in order to simplify Maxwell’s equations for a system moving uniformly with velocity \(v\) relative to the ether.  
To first order in \(v/c\), he introduced the transformation
\[
x' = x - vt, \qquad y' = y, \qquad z' = z, \qquad t' = t - \frac{v}{c^{2}}\,x .
\]

The new time variable \(t'\), which Lorentz called \emph{local time}, was not a mere mathematical symbol.  
In 1895 it still functioned primarily as an auxiliary time used in the theorem of corresponding states, while Lorentz retained a true time \(t\) associated with the ether.  
Yet the variable already had a clear physical role: it allowed optical and electromagnetic phenomena in a moving system to be described as if that system were at rest.  
The local time was therefore not yet the universal time of an inertial frame in Einstein’s sense, but it was already tied to the physical explanation of the apparent independence of optical phenomena from the Earth’s motion.

This step was decisive.  
Lorentz gave priority to electrodynamics over ordinary Galilean kinematics and introduced a transformation of time as well as of space.  
Even though he did not yet abandon the ether frame or the distinction between true time and local time, the 1895 theory contained one of the central mathematical and physical elements from which the later theory of relativity would emerge.

The 1895 theory also had limits.  
It accounted for many first-order optical effects, but it did not yet provide by itself a complete explanation of the Michelson--Morley experiment, which required second-order effects in \(v/c\).  
To explain that result, Lorentz had to rely on the contraction hypothesis, first introduced by FitzGerald and Lorentz in the early 1890s.  
Thus, by 1895, Lorentz had already assembled two essential ingredients of the later theory: local time and length contraction, but they remained embedded in an ether-based electron theory.

\subsection{Lorentz’s Electron Theory and the 1904 Amsterdam Memoir}

Lorentz’s work around 1904 must be understood against the background of his broader electron theory.  
In the same period, he published a long synthetic article in the \emph{Enzyklopädie der mathematischen Wissenschaften mit ihren Anwendungen}.  
This article presented the development of Maxwell’s theory into the electron theory and offered a systematic account of Lorentz’s programme.  
It was a major reference work for the German-speaking scientific community and gave Lorentz’s theory considerable visibility.

The specific paper that became central to the later history of special relativity was, however, a shorter memoir published in the \emph{Proceedings of the Royal Academy of Sciences of Amsterdam}: \emph{Electromagnetic Phenomena in a System Moving with Any Velocity Smaller than That of Light} \cite{lorentz1904}.  
This paper appeared in 1904 and addressed the problem of extending Lorentz’s theory beyond first order.  
Its aim was to account for the negative results of ether-drift experiments, including Michelson--Morley and Trouton--Noble, by constructing a transformation capable of preserving the relevant form of the electromagnetic equations for a system in uniform motion.

\subsection{Poincaré’s Criticism and Lorentz’s 1904 Transformation}

Around 1900, Henri Poincaré emerged as one of the most perceptive and demanding commentators on Lorentz’s electron theory.  
In \emph{La théorie de Maxwell et les oscillations hertziennes} and in several papers, he highlighted the difficulties of Lorentz’s model.  
He pointed out, in particular, the conceptual difficulties raised by the hypothesis of an immobile ether.  
In his 1900 paper on Lorentz’s theory and the principle of reaction, he also gave Lorentz’s local time a clearer physical interpretation in terms of clock synchronization by light signals.

More generally, Poincaré stressed the need to formulate the principle of relativity with greater rigor and to clarify the transformation laws of electrodynamics.

Lorentz’s 1904 paper can be read as an attempt to strengthen his theory in response to precisely these difficulties.  
He extended his earlier first-order transformation and introduced a more general transformation involving the factor now denoted by \(\gamma\), together with an undetermined scale factor \(l(v)\).  
For uniform motion with velocity \(v\) along the \(x\)-axis, the transformation may be written in the form:
\[
x' = l(v)\,\gamma(x - vt), \qquad 
y' = l(v)\,y, \qquad 
z' = l(v)\,z, \qquad 
t' = l(v)\,\gamma\!\left(t - \frac{vx}{c^{2}}\right),
\]
where
\[
\gamma = \frac{1}{\sqrt{1-v^{2}/c^{2}}}.
\]
The presence of \(\gamma\) shows that the speed of light had become structurally central to the transformation.  
The factor \(l(v)\), however, remained undetermined in Lorentz’s treatment; it would be fixed by Poincaré in 1905 through the group property of the transformations.

In 1904, the status of \(t'\) became stronger than in 1895.  
Lorentz used local time in the construction of corresponding states and connected it with the behavior of electromagnetic systems in motion.  
In particular, his treatment of moving electromagnetic oscillators implied a real retardation of their periods.  
Thus \(t'\) was not simply a formal variable: it was tied to the time indicated by ideal electromagnetic clocks in the moving system.  
Nevertheless, Lorentz continued to distinguish this local time from the true time of the ether frame.

Lorentz interpreted the empirical success of his theory through two main mechanisms:
\begin{enumerate}
    \item \textbf{Length contraction.}  
    Bodies moving through the ether were assumed to contract in the direction of motion.  
    This contraction was not a purely kinematical consequence of the relativity principle, but a dynamical hypothesis concerning the constitution of matter and the electromagnetic forces binding it.

    \item \textbf{Retardation of electromagnetic processes.}  
    Lorentz showed that electromagnetic systems serving as clocks, such as idealized oscillators, would have their periods altered when in motion.  
    This result gave local time a direct physical relevance: it was connected with the time actually indicated by such moving electromagnetic clocks.  
    But Lorentz did not transform this result into a universal kinematical principle applying to all possible clocks independently of their physical constitution.
\end{enumerate}

Thus, the negative results of ether-drift experiments received a systematic explanation within Lorentz’s framework.  
Length contraction and the modification of electromagnetic processes made it impossible, within the theory, to detect the Earth’s motion through the ether by the experiments then under discussion.  
But the interpretation remained Lorentzian: the ether frame still existed, true time was still distinguished from local time, and the principle of relativity was not yet elevated to the status of a universal postulate.

\subsection{Immediate Reception of Lorentz’s Work in Germany and France}

The reception of Lorentz’s work was particularly swift in German-speaking circles. 
In July 1904, Wilhelm Wien published a short paper discussing the foundations of electrodynamics for moving bodies and engaging directly with Lorentz’s results \cite{wien1904}. Its importance lies not only in the fact that Wien discussed Lorentz’s theory, but also in the extreme visibility of the transformation itself: in a text of only a few pages, Wien explicitly wrote down the Lorentz transformation.  
Soon afterward, Wien replied to criticism by Max Abraham, confirming that Lorentz’s theory had already become part of an active German debate. 

At the same time, Max Abraham defended his own theory of the rigid electron.  He acknowledged the importance of Lorentz’s work while rejecting the contraction hypothesis and the use of local time as Lorentz employed it.  
Abraham’s opposition is important: it shows that Lorentz’s theory was not ignored, but immediately became a point of reference even for those who rejected its premises.

Emil Cohn also entered the discussion in 1904 with his own electrodynamics of moving systems \cite{cohn1904}. His two communications proposed an alternative formulation of electrodynamics for moving bodies. Cohn’s approach differed from Lorentz’s, but it was explicitly situated in relation to the Lorentzian programme.  
Here again, the important point is not agreement but visibility: Lorentz’s theory had become unavoidable in the German debate on electrodynamics.

Sommerfeld’s work on electron theory provides another indication of this rapid reception. In 1904 and 1905, he published papers on the dynamics and fields of electrons, drawing on problems opened by Lorentz’s electron theory and by Abraham’s competing model. These publications show that the Lorentzian programme was being actively discussed and technically developed by some of the leading younger theoretical physicists in Germany.

Lorentz’s broader electron theory also circulated through the \emph{Enzyklopädie der mathematischen Wissenschaften mit ihren Anwendungen}, one of the most prestigious vehicles of mathematical and physical knowledge in the German-speaking world.  
His article was not a translation of the Amsterdam memoir on electromagnetic phenomena in moving systems, but a large synthetic presentation of the electron theory. Together with the Amsterdam paper of 1904, it helped make Lorentz’s programme highly visible.

The rapid publications by Wien, Abraham, Cohn, and Sommerfeld show that Lorentz’s ideas were immediately known and debated in the main German-speaking theoretical physics circles. The diffusion of Lorentz’s concepts was thus well established by late 1904 and early 1905 within a prestigious editorial and institutional context.

At the International Congress of Arts and Science held in St. Louis in September 1904, references to Lorentz’s work were also made by Henri Poincaré and Paul Langevin.  
In his lecture \emph{The Present and Future of Mathematical Physics}, Poincaré presented Lorentz’s theory as the most advanced attempt to address the electrodynamics of moving bodies.  
As will be seen later, this lecture was quickly published through several channels. Langevin likewise discussed the recent development of electron theory and treated Lorentz’s methods as central to the problem of reconciling optics, electrodynamics, and the failure to detect the Earth’s motion through the ether.

All these reactions illustrate the rapid and transnational diffusion of Lorentz’s ideas from 1904 onward. Within a few months, his work was being discussed in major physics journals, academies, encyclopedic publications, and international congresses.  
This immediate resonance shows that Lorentz’s results already stood as a central reference for analyzing the electrodynamics of moving bodies.

\section{Richard Gans’s Review: A Gateway to Lorentz’s 1904 Paper}

\subsection{Editorial Context}

As already noted, at the beginning of the twentieth century, the \emph{Beiblätter zu den Annalen der Physik} was a review and bibliographical journal that played a key role in the rapid dissemination of scientific advances. Its readership naturally overlapped with that of the \emph{Annalen}, but its function was different: it allowed physicists to follow a rapidly expanding literature without having to consult every original article.  
Within this editorial landscape, a review published in the \emph{Beiblätter} could make an important foreign publication quickly visible to the German-speaking scientific community.

\subsection{Richard Gans: A Recognized Mediator}

Richard Gans (1880--1954) was a young German theoretical physicist trained under Paul Drude and Max Abraham, soon to become known for his contributions to the electrodynamics of matter and to optics. In the early 1900s, he was already a well-informed specialist in electromagnetism and in the propagation of light in metals. His review of Lorentz’s 1904 paper therefore carried a double authority:
\begin{itemize}
  \item that of the \emph{Beiblätter} itself, an important medium of scientific dissemination;  
  \item and that of Gans, a competent reader of Maxwellian electrodynamics, Lorentz’s theory, and the contemporary debates on moving bodies.  
\end{itemize}
This context reinforces the weight of his report: it was a substantial review written by a physicist familiar with the field.

\subsection{Richard Gans’s 1905 Review}

The text is remarkably clear and well written \cite{gans1905}. In a few pages, Gans highlighted the key points of Lorentz’s 1904 memoir without becoming lost in technical details. He began by recalling three experimental difficulties. First, the original Lorentz electron theory did not explain why the Earth’s motion had no observable influence on the interference of light in the Michelson--Morley experiment.  
Second, it did not explain why no torque acted on a charged plane capacitor in the Trouton--Noble experiment.  
Third, the FitzGerald--Lorentz contraction hypothesis, introduced to account for Michelson--Morley, seemed to require a double refraction of light in isotropic bodies due to the Earth’s motion; the experiments of Lord Rayleigh and Brace had again yielded negative results.

Gans then summarized Lorentz’s answer to these difficulties.  
By introducing a transformation of space and the local time \(t'\), Lorentz obtained equations in the transformed moving system that had exactly the same form as the Lorentz equations in the original system at rest.  
Gans emphasized that this correspondence held rigorously, not merely to second order.  
In electrostatic and optical fields, no effect of any order of the motion could be detected.

The review also mentioned the transformation of ponderomotive forces per unit volume and emphasized that Lorentz extended the same rule to non-electromagnetic forces, such as elastic forces.  
From this principle followed the idea that a body held in equilibrium by its internal attractions and repulsions would, by its motion, change its dimensions in the required way.  
Thus the contraction of bodies was no longer presented simply as an isolated hypothesis, but as a consequence of a broader transformation rule applied to the forces maintaining equilibrium.

Through its concision and precision, Gans’s review made accessible, in German, the essential elements of Lorentz’s 1904 memoir: the transformation of space and time coordinates, the use of local time, the contraction of electrons and material bodies, the transformation of forces, and the explanation of the negative results of Michelson--Morley, Trouton--Noble, and Rayleigh--Brace.  
Its importance is therefore considerable.  
At the beginning of 1905, a German-speaking physicist attentive to the \emph{Beiblätter} could obtain from this notice a clear summary of Lorentz’s 1904 results.

\subsection{A Text Difficult to Ignore}

Under the scientific conditions of 1905, this was not an obscure or inaccessible text.  
A physicist regularly following the \emph{Beiblätter} would have encountered a concise summary of Lorentz’s latest results in a journal specifically designed to report recent international literature.  
Several features made the notice especially visible:
\begin{itemize}
  \item the \emph{Beiblätter} were addressed precisely to readers wishing to follow the current literature of physics;  
  \item Gans’s review condensed Lorentz’s innovations into a short and readable text;  
  \item it clearly identified the central issues: the transformation of space and time, the use of local time, the contraction hypothesis, the transformation of forces, and the explanation of the negative results of Michelson--Morley, Trouton--Noble, and Rayleigh--Brace.  
\end{itemize}

Gans’s review helps explain how Lorentz’s ideas could circulate rapidly and become known in German-speaking circles by the beginning of 1905.  
Any subsequent discussion of relativity took place within an intellectual environment where Lorentz’s results were already disseminated through several channels: the Amsterdam memoir, discussions in German physics journals, Lorentz’s broader electron-theory article in the \emph{Enzyklopädie}, and this concise review in the \emph{Beiblätter}.

Yet this document has received limited attention from historians.  
Most standard accounts of the genesis of special relativity either ignore it or mention it only briefly, without analyzing its role as a channel of dissemination.  
Michel Janssen is one of the few historians to have emphasized its importance, though only briefly, in his 1995 doctoral dissertation \cite{janssen1995}.  
To my knowledge, one of the rare published discussions of the possible role of Gans's review appears in C.~Bracco and J.~Provost \cite{bracco-provost2018}.  
They note that Lorentz's new expression for local time was reproduced in the \emph{Beiblätter} in the second half of February 1905 and argue that, since Einstein was then serving as a reviewer for the journal, he must have become acquainted with it no later than March 1905.

\section{Henri Poincaré’s Work on Relativity, 1898--1906}

\subsection{1898: The Operational Definition of Time}

In 1898, in his article \textit{La mesure du temps} \cite{poincare1898}, Poincaré approached the question of time through the concrete problem of measurement.  
How can two distant clocks be synchronized?  
His answer was operational: one may synchronize them by means of light signals, on the convention that light takes the same time to travel in the two opposite directions.  
Simultaneity is therefore not simply given by nature; it is defined through a practical rule involving clocks, signals, and an assumption about the propagation of light.

This analysis is one of Poincaré’s most important anticipations of the later conceptual structure of relativity. It introduced a decisive idea: the time used in physics is inseparable from procedures of measurement and synchronization.  
In this sense, Poincaré had already opened the way to an operational analysis of temporal concepts.

\subsection{1900: Local Time, Action--Reaction, and Electromagnetic Momentum}

Poincaré’s 1900 paper \textit{La théorie de Lorentz et le principe de réaction} \cite{poincare1900} is one of the most important texts in the prehistory of special relativity.  
It was published in the Lorentz jubilee volume, in the \emph{Archives néerlandaises des sciences exactes et naturelles}, and was devoted to a difficulty that Poincaré regarded as fundamental: the apparent violation of the principle of action and reaction in Lorentz’s electron theory.
  
For Poincaré, the principle of reaction belonged to the highest level of physical theory, because it expressed the conservation of the motion of the center of gravity of an isolated system. Lorentz’s theory, however, seemed to violate this principle.  
In the interaction between matter and ether, action and reaction were not balanced in the ordinary mechanical sense: matter acted on the ether, and the ether acted on matter, but the total mechanical balance was obscure. Poincaré therefore saw in Lorentz’s theory a deep tension between electrodynamics and the principles inherited from mechanics.

As Olivier Darrigol has emphasized \cite{Darrigol2023}, this concern with the reaction principle has often been neglected or underestimated in historical accounts of relativity, which tend to focus almost exclusively on the principle of relativity.  
Yet for Poincaré the two issues were closely connected. The electrodynamics of moving bodies had to satisfy not only the requirement that absolute motion remain undetectable, but also the requirement that the conservation principles of mechanics be preserved or reformulated in a coherent way.  
The problem of relativity was therefore, for Poincaré, inseparable from the problem of momentum, reaction, and the dynamical role of the electromagnetic field.

In the same paper, Poincaré gave a physical interpretation of Lorentz’s local time. He explained that two observers moving with the Earth could synchronize their clocks by exchanging light signals, assuming that light travels with the same speed in both directions.  
Because the observers are moving through the ether, this assumption is not true from the standpoint of the ether frame; nevertheless, it leads them to adopt precisely Lorentz’s local time. Thus, already in 1900, local time was no longer merely an algebraic device. Poincaré interpreted it as the apparent time indicated by clocks synchronized by light signals in a moving system.

This interpretation is very important.  It shows that Poincaré had understood the physical meaning of Lorentz’s local time before Einstein’s 1905 paper.  
The difference is not that Lorentz and Poincaré had only a meaningless mathematical variable whereas Einstein alone gave it physical content.  
Rather, the difference is that Poincaré still maintained a distinction between true time, attached to the ether frame, and local or apparent time, attached to observers moving through the ether. 

Poincaré’s 1900 paper also introduced another decisive idea: electromagnetic radiation carries momentum. To preserve the center-of-gravity theorem, he associated an effective momentum with electromagnetic energy, and introduced what he called a fictitious fluid of energy moving with the velocity of light. This led him to the relation between electromagnetic energy and an effective mass, later explicitly acknowledged by Einstein in his 1906 paper on the inertia of energy. Thus, Poincaré 1900 is important not only for local time, but also for the emergence of the connection between energy, momentum, and inertia.

Darrigol’s analysis is particularly valuable here because it restores the internal logic of Poincaré’s argument.  
Poincaré was not merely accumulating isolated anticipations of later relativity.  
He was trying to repair the foundations of electrodynamics by imposing two high-level constraints: the principle of relative motion and the principle of reaction.  
His 1900 paper therefore marks a decisive stage in the transition from Lorentz’s electron theory to a more general relativistic structure.

\subsection{The Book \emph{La Science et l’hypothèse}}

Published in 1902, \emph{La Science et l’hypothèse} is Poincaré’s most influential philosophical work.  
Developed from lectures and essays written in the preceding years, it presented to a broad audience his reflections on the foundations of science.  
Translated rapidly into several languages---including German in 1904---the book achieved wide international circulation.  
In the German translation, the translators added a footnote referring explicitly to Poincaré’s article \textit{La mesure du temps}, thereby giving German-speaking readers a direct indication of a text absent from the original French volume.

The book contains no technical development of the Lorentz transformations, but it offers a strikingly clear philosophy of physical science.  
Poincaré argued that absolute space has no empirical reality: only relations among bodies, and the measurement of distances and motions, have physical meaning.  
Likewise, absolute time is not directly given by experience.  
The simultaneity of two distant events can be defined only through a convention, for example by assuming that light covers the two directions of a round trip in equal times.

Poincaré also discussed what he called the principle of relative motion.  
The known laws suggested that no mechanical, optical, or electrical experiment could reveal absolute motion.  
He reviewed experimental attempts, especially optical and interferometric ones, that had failed to detect the Earth’s motion through the ether.  
In this philosophical context, the principle of relativity appeared not as a sudden isolated postulate, but as the gradual elevation of a repeated empirical failure into a general guiding principle.

More generally, Poincaré emphasized the hypothetical and provisional nature of scientific laws.  
Theories are not final copies of reality, but instruments for classification, prediction, and coordination of experience.  

\subsection{1904: The Principle of Relativity at the St. Louis Lecture}

In 1904, at the St. Louis Exposition, Poincaré took a further step.  
He proposed a general principle:
\begin{quote}
The laws of physical phenomena must be the same for an observer fixed on the Earth and for an observer carried along in a uniform motion of translation; so that we have not, and cannot have, any means of discerning whether or not we are carried along in such a motion.
\end{quote}
The lecture delivered by Poincaré in September 1904, entitled \textit{L’état actuel et l’avenir de la physique mathématique}, circulated widely and rapidly.  
The full text appeared in \emph{La Revue des Idées} on 15 November 1904 \cite{poincare1904revue-idees} and, a month later, in the \emph{Bulletin des sciences mathématiques} in December 1904 \cite{poincare1904}.  
In January 1905, a complete English translation appeared in \emph{The Monist} under the title \textit{The Principles of Mathematical Physics} \cite{poincare1905-monist}.  
The lecture was also published in volume~1 of the congress proceedings, \emph{Actes du Congrès international des arts et des sciences, Exposition universelle de Saint-Louis, 1904}.

The lecture was then reprinted, almost verbatim, at the opening of \emph{La Valeur de la science} in 1905 \cite{poincare1905-valeur}.  
Through these editorial channels Poincaré ensured rapid international circulation of his exposition, which contains one of the clearest pre-1905 statements of the principle of relativity.

The impact was immediate.  
In Göttingen, Poincaré’s St. Louis lecture was discussed in the Mathematical Society beginning on 31 January 1905, and the discussion continued at the next meeting on 7 February.  
The address opened the 1905 series of discussions devoted to electron theories and the foundations of relativity.  
Among the participants were several leading figures of German mathematics and physics, including David Hilbert, Felix Klein, Hermann Minkowski, and probably Max Born and Max Laue.

This sequence attests to the rapid dissemination of Poincaré’s lecture.  
Published in France and the United States, placed at the head of a successful book, and debated in one of the foremost centers of mathematical and physical research, it helped establish the principle of relativity at the heart of scientific debates from the beginning of 1905.

Poincaré presented the principle of relativity as a general guide for rewriting physics as a whole, not merely electrodynamics. He treated the ether as dynamically undetectable and increasingly deprived of empirical function, while still maintaining the distinction between true time and local time.

An important testimony was later provided by Lorentz in 1921, in his article \textit{Deux mémoires de Henri Poincaré sur la physique mathématique}, published in \emph{Acta Mathematica}.  
Reviewing the history of electrodynamic theories, Lorentz acknowledged that his own work had not elevated the principle of relativity to the status of a universal law. By contrast, he emphasized that Poincaré had explicitly formulated what Lorentz called the postulate of relativity, an expression whose introduction Lorentz attributed to Poincaré.

\subsection{1905: The Lorentz Group and Covariance}

In his 1905 note \cite{poincare1905}, Poincaré once again placed the problem within the framework of the principle of relativity.  
He began from the repeated failure of optical and electromagnetic experiments, including Michelson's, to reveal the absolute motion of the Earth, and described Lorentz's theory as an attempt to bring the electrodynamics of moving bodies into agreement with the postulate of the complete impossibility of determining such motion.

Poincaré then made a decisive mathematical step.  
He showed that the Lorentz transformations must form a group.  
They compose with one another, possess inverses, and include the identity transformation.  
From this group property, Poincaré determined the function \(l(v)\), left undetermined by Lorentz in 1904, and concluded that it must be equal to 1.

He also established the invariance of the equations of the electromagnetic field under the Lorentz transformation and derived the corresponding transformation laws for the electric density, the velocity of the charge, and the electromagnetic force, both per unit volume and per unit mass of the electron.  
It is in the transformation formulae for the force that Poincaré noted differences from Lorentz's expressions; in particular, one complementary term involving the scalar product of force and velocity.

The note also extended the question beyond pure electrodynamics.  
Poincaré discussed the contraction hypothesis for the electron, compared the assumptions of Lorentz and Langevin, and introduced a compensating pressure capable of accounting for the contracted form of the electron.  
He then examined how the law of gravitation would have to be modified in order to be compatible with the Lorentz transformation.  
Already in the short note, he proposed that gravitation should propagate with the speed of light and indicated the form of a corrected attraction law; the fuller treatment would be developed in the Palermo memoir.
Poincaré thus elevated the Lorentz transformation to the status of an organizing principle of physics.  
His contribution was not merely to correct a detail in Lorentz's theory, but to recognize the group-theoretical structure underlying the new electrodynamics and to use the principle of relativity as a criterion for admissible physical laws.

\subsection{Poincaré 1906: Systematizing the Principle of Relativity and the Role of the Principle of Least Action}

In his major Palermo memoir, submitted in July 1905 and published in the 1906 volume of the \emph{Rendiconti} \cite{poincare1906}, Poincaré gave the complete technical and mathematical elaboration of the results announced in his brief 1905 note.  
He systematically developed the Lorentz transformations, their group structure, and their consequences for electrodynamics, electron dynamics, and gravitation.  
This work contains several elements of great importance:
\begin{itemize}
\item the explicit and detailed demonstration that the Lorentz transformations form a group, which Poincaré called the Lorentz group;
\item the law of velocity addition;
\item the transformation laws for electric density, charge velocity, current, electromagnetic fields, potentials, and force;
\item the rewriting of Lorentz's electrodynamics in a covariant form;
\item the use of a four-dimensional formal analogy, especially through the invariant quadratic form;
\item the integration of the principle of least action into the Lorentz-invariant formulation of electron dynamics.
\end{itemize}

The last point is particularly important.  
Poincaré did not merely verify the covariance of particular equations.  
He showed that the principle of least action could be used to organize Lorentz's theory and to explain why the Lorentz transformation preserves the form of the electromagnetic laws.  
This method did not yet amount to the later Minkowskian geometry of space-time, nor to the modern language of relativistic field theory.  
But it anticipated a style of theoretical physics in which invariance principles and variational formulations play a central role.

Poincaré's Palermo memoir is therefore the most complete pre-Minkowskian systematization of the principle of relativity.  
It retains the ether and does not abandon the distinction between true and local time, but it organizes physical laws according to Lorentz covariance and gives the theory a powerful mathematical structure.

Despite this powerful mathematical systematization, Poincaré surprisingly did not recall here the physical interpretation of local time that he had already presented in 1900, in terms of the synchronization of moving clocks by light signals. 

\section{Challenges to the Ether}

At the turn of the twentieth century, several scientists began to question the ether, or at least to deprive it of any detectable mechanical function.  
The repeated failure to observe the Earth’s motion through the ether did not immediately lead to its abandonment, but it transformed its status.  
For some physicists, the ether remained a useful theoretical support for electromagnetic waves; for others, it increasingly appeared as an undetectable remnant of an older mechanical picture of nature.

Ernst Mach played an important role in this broader intellectual climate.  
His criticism was not directed only, or even primarily, at the electromagnetic ether, but at all concepts that claimed physical significance without clear empirical anchoring.  
In \emph{The Science of Mechanics: A Critical and Historical Account of Its Development}, he criticized absolute space and absolute motion and insisted that mechanics should be reconstructed from observable relations among bodies.  
This anti-metaphysical stance encouraged skepticism toward entities that could not be connected to possible experience.

In later writings, Mach continued to emphasize the economy of thought as a methodological rule.  
Scientific concepts should simplify and coordinate experience; they should not multiply invisible mechanisms without increasing the predictive or explanatory power of theory.  
In his 1903 lecture \emph{Space and Geometry from the Point of View of Physical Inquiry}, he again tied spatial and geometrical concepts to physical inquiry and measurement.  
Within such a framework, a mechanical ether that produced no observable effects could easily appear as a dispensable hypothesis rather than as a necessary element of physics.

This attitude helped prepare a philosophical climate in which the ether could be treated with suspicion. If a supposed medium could not be detected and did not improve the empirical content of the theory, its physical status became increasingly problematic.

One may also mention Emil Cohn, a German theoretical physicist who, between 1900 and 1904, proposed an alternative electrodynamics of moving bodies.  
Cohn did not simply reproduce Lorentz’s electron theory; he sought to formulate electrodynamics in a way that reduced, or even removed, the role of the traditional ether.  
His work shows that the ether was already being questioned not only on philosophical grounds, as in Mach, but also within technical electrodynamics itself.

\subsection{Poincaré and the Ether}

From the turn of the century onward, Poincaré spoke on several occasions about the status of the ether, clearly relativizing its importance.  
He did not simply reject it but he repeatedly stressed that its existence had become difficult to distinguish from a convention.

In \emph{La Science et l’hypothèse}, Poincaré emphasized that the existence or nonexistence of the ether made little difference to actual physics, insofar as no experiment could reveal the Earth’s motion with respect to it.  
The ether thus became, in his hands, less a directly observable substance than a convenient theoretical support.  
In his 1904 St. Louis lecture, he conceded that ``perhaps the ether will be rejected as useless,'' although he still preferred to retain it provisionally.  
Finally, in \emph{La Valeur de la science} (1905), he again treated the ether as a hypothesis that physics might eventually be able to do without.

These statements show that, for Poincaré, the ether was not fundamental in the same sense as the principle of relativity or the covariance of physical laws.  
It remained a conceptual support, useful but contingent.  
Its eventual disappearance would not, in his view, destroy the coherence of physics, provided that the laws themselves retained the proper invariant structure.  
In this respect, Poincaré’s position occupied an intermediate place: he did not abolish the ether but he had already emptied it of much of its empirical and mechanical content.

\section{Einstein’s Scientific Readings Between Graduation and His Arrival in Bern}

\subsection{The Library of the Istituto Lombardo and Einstein’s Readings in Milan}

This section draws on Christian Bracco’s studies of Einstein’s stays in Milan \cite{Bracco2015, Bracco2016}.  
At the turn of the century, the library of the Istituto Lombardo, located in the Palazzo Brera in Milan, was one of Italy’s important documentary centers in the exact sciences.  
Its holdings, built up through exchanges with leading academies and scientific institutions, included major journals of physics and mathematics, as well as collective volumes and recent scientific reports.  
This bibliographical wealth made it an important resource for a young researcher wishing to keep up with recent developments in physics.

Einstein’s letters to Mileva Marić indicate that he used a scientific library during his stays in Milan between 1899 and 1901.  
Bracco has argued that this library can be identified with the library of the Istituto Lombardo, since it possessed the journals Einstein mentions or appears to have consulted, including the \emph{Annalen der Physik} and probably the \emph{Beiblätter zu den Annalen der Physik}.  
Living with his family at 21 Via Bigli, only a short distance from the Palazzo Brera, Einstein could consult journals in which the works of Planck, Wien, Drude, Lorentz, Boltzmann, Poincaré, and others were published.  
This access allowed him to supplement the education he had received at the Zurich Polytechnic.  
The Istituto Lombardo thus offered him a direct opening onto international scientific literature.  

The importance of this access extended beyond the university holidays.  
After graduating in July 1900 and before obtaining his first temporary teaching posts in May 1901, Einstein continued to reside for periods in Milan and to work on scientific problems there.  
One particularly important item was the \emph{Festschrift} offered to Lorentz in December 1900 for the twenty-fifth anniversary of his doctorate.  
This volume gathered contributions from many leading physicists, including Boltzmann, Planck, Poincaré, Wien, Zeeman, and Lorentz himself.  
Bracco has shown that the volume was received by the library of the Istituto Lombardo on 31 January 1901, as indicated by an inscription on its cover.  
Since Einstein was then engaged in bibliographical research for his doctoral work, it is highly plausible that he could have consulted this volume in Milan in the spring of 1901.

The \emph{Festschrift} provided a condensed view of contemporary debates in electrodynamics, kinetic theory, and molecular physics.  
Lorentz did not merely receive the commemorative volume; he also contributed to it with an important paper connected with his earlier work on the electrodynamics of moving bodies and the electron theory.  
There, Lorentz developed an analysis of electromagnetic fields in systems in uniform motion, seeking to account for observed optical phenomena.  
One already finds in this text several conceptual premises that would lead to the more complete transformation theory of 1904.

Thus, the Istituto Lombardo library served as an important intellectual relay between Zurich and Milan, granting Einstein access to a European scientific literature that was important for his formation.  
This context matters because it shows that, even before Bern, Einstein was not intellectually isolated.  
He already knew how to use libraries, correspondence, and personal networks to obtain recent scientific information.

\subsection{Einstein’s Letter to Mileva Marić of 15 April 1901}

On 15 April 1901, writing from Milan, Einstein asked Mileva Marić, who had remained in Zurich, to send him Kirchhoff’s treatise on heat by mail.  
This request is revealing.  
It shows, first, that despite his access to the library of the Istituto Lombardo, Einstein could not always find locally every work he needed.  
The letter also sheds light on Einstein’s research practices at this stage of his career.  
Rather than limiting himself to local resources, he mobilized his personal network: Mileva acted as an intermediary, able to forward him books available in Zurich.  
Correspondence thus complemented library access.  
This method illustrates the concrete working conditions of a young physicist at the turn of the century: books and articles circulated between cities, and personal relations could play an essential role in overcoming practical obstacles.

\subsection{Einstein’s Early Interest in Lorentz’s Work: The Letter of 28 December 1901 to Mileva Marić}

At the end of 1901, Albert Einstein, aged twenty-two, was living in Schaffhausen, a town in northern Switzerland about fifty kilometers from Zurich, near the German border.  
He had obtained a position in a private school after a temporary teaching appointment in Winterthur.  
His task was to prepare a young student, who intended to study at the Zurich Polytechnic, for the Matura examination.

In a letter dated 28 December 1901, Einstein told Mileva Marić that he had asked his former classmate Jakob Ehrat to send him the recent works of Lorentz.  
Ehrat had studied at the Zurich Polytechnic in the same cohort as Einstein and, after graduating in 1900, remained there as an assistant.  
From this position, he had easier access to libraries and specialized journals than his friend, who was then relatively isolated in Schaffhausen.

A few weeks later, in February 1902, a dispute with his director brought Einstein’s employment to an early end.  
He passed through Zurich to withdraw his doctoral dissertation, which had not been accepted, and then moved to Bern on 3 February 1902.  
There, he sought a position at the Federal Patent Office, an appointment that would not be officially confirmed until 23 June 1902.

The December 1901 letter is highly significant.  
It shows that Einstein, at that date, was already actively seeking to consult recent publications by Lorentz.  
To do so, he relied on his personal network, asking a former classmate to send him the materials by post.

Two main conclusions follow:
\begin{itemize}
  \item Einstein’s interest in Lorentz’s electrodynamics was early, explicit, and concrete.
  \item The circulation of documents played an essential role: when direct access to specialized libraries was limited, Einstein used personal connections to obtain recent works.
\end{itemize}

Despite the vast historiography devoted to the genesis of special relativity, this request to Jakob Ehrat has received little attention.  
The editors of the \emph{Collected Papers} note the passage in the letter, but the major historical accounts of the origins of relativity rarely make it central to their analysis.  
Yet it is a valuable piece of evidence: before his arrival in Bern, Einstein was already actively seeking Lorentz’s recent work and was accustomed to obtaining scientific literature through a combination of libraries, correspondence, and personal networks.

\section{Einstein’s 1905 Paper on Relativity}

\subsection{Einstein and Special Relativity (1905)}

The number of analyses devoted to this work is immense, so only a brief outline will be given here.  
Einstein based his reasoning on two postulates: the principle of relativity and the constancy of the speed of light.  
He then defined an operational procedure for synchronizing clocks using light signals \cite{einstein1905}.  
This was one of the central conceptual moves of the paper: time was no longer treated as an immediately given universal parameter, but as a quantity defined through physical operations involving clocks and light signals.

Einstein’s derivation of the transformation between two inertial systems also relied on general assumptions that are now commonly described in terms of homogeneity and isotropy.  
The homogeneity of space and time implies that the transformation between coordinates must be linear, while the isotropy of space restricts the possible dependence on the direction of motion.  
Einstein considered two systems \(S\) and \(S'\), the latter moving uniformly relative to the former along the \(x\)-axis, and sought the relation between the coordinates \((t,x,y,z)\) and \((t',x',y',z')\) of the same event.  
The coefficients of this transformation could depend on the relative velocity \(v\), while the transverse coordinates were constrained by symmetry.

\subsection{Constraints on the Transformation}

This transformation is then constrained by several arguments.  
The principle of relativity imposes reciprocity: the inverse transformation must have the same form when the relative velocity is reversed.  
The constancy of the speed of light imposes the condition that light signals satisfying \(x=\pm ct\) in \(S\) must also satisfy \(x'=\pm ct'\) in \(S'\).  
Together with the symmetry assumptions, these requirements lead to the Lorentz transformation, up to a factor that Einstein eliminates by an additional symmetry argument.

The result is formally identical to the transformation previously obtained by Lorentz within electron theory and by Poincaré through the analysis of invariance and group structure.  
Einstein’s derivation, however, gave it a different foundation.  
Instead of deriving the transformation from the detailed structure of Maxwell’s equations or from a theory of electrons moving through the ether, he derived it from two general postulates concerning inertial frames and the propagation of light.  
This shift in foundation is one of the distinctive features of the 1905 paper.

\subsection{The Addition of Velocities}

On this basis, Einstein derived the law of \emph{velocity addition} consistent with the constancy of \(c\).  
For motion along the common \(x\)-axis of the two frames, if a particle has velocity \(u_x\) in \(S\), its longitudinal velocity in \(S'\) is
\[
u'_x = \frac{u_x - v}{\,1 - \dfrac{u_x v}{c^{2}}\,}.
\]
The transverse components transform differently:
\[
u'_y = \frac{u_y \sqrt{1-v^2/c^2}}{\,1 - \dfrac{u_x v}{c^{2}}\,},
\qquad
u'_z = \frac{u_z \sqrt{1-v^2/c^2}}{\,1 - \dfrac{u_x v}{c^{2}}\,}.
\]
These formulae replace the Galilean addition of velocities and ensure that the composition of subluminal velocities remains subluminal.  
They also show that \(c\) is invariant under the transformation: a light signal moving with velocity \(c\) in one inertial frame also has velocity \(c\) in the other.

\subsection{Application to Electrodynamics}

Einstein then applied these results to electrodynamics.  
He showed that the Maxwell--Lorentz equations retain their form under the new transformation, and that the electric and magnetic fields transform into one another according to the state of motion of the observer.  

The paper contains no formal bibliography. It does, however, mention several names in the body of the text, including Maxwell, Hertz, and Doppler.  
Lorentz’s 1895 and 1904 memoirs and Poincaré’s works on the principle of relativity are not cited.  
It opens with the well-known asymmetry between the two usual descriptions of electromagnetic induction: either a magnet moves near a stationary conductor, or a conductor moves in the field of a stationary magnet. Physically, only the relative motion matters, since the observable effect is the induced current.  
Yet the traditional theoretical descriptions were different. When the magnet is moved, one invokes an electric field produced in the surrounding space; when the conductor moves, one describes the magnetic force acting on charges in the conductor. Einstein treated this asymmetry as a sign that the distinction between the two descriptions was artificial. Together with the failure to detect motion relative to the ether, it pointed toward a formulation in which the laws of electrodynamics would have the same form in all inertial frames. This led him to elevate the principle of relativity and the constancy of the speed of light into the two starting points of the theory.

Einstein also mentioned two long-known optical phenomena: stellar aberration, discovered by Bradley in the eighteenth century, and the Doppler effect, identified in the nineteenth century.  
By contrast, he passed over in silence almost the entire recent background of the problem: the ether-drift experiments, Lorentz's 1904 theory, Poincaré's formulations of the principle of relativity, and the broader electrodynamical debate of the previous two decades.  
This silence is one of the most striking features of the 1905 paper, whose argumentative force depends in part on its radical decontextualization of the problem.

\subsection{Michele Besso}

At the end of the paper, Einstein explicitly thanked Michele Besso, acknowledging the ``faithful assistance'' of his friend and colleague at the Patent Office and ``several valuable suggestions.''  
For more information on Michele Besso and his role in Einstein’s life, one may consult the correspondence edited by Pierre Speziali (\emph{Einstein--Besso Correspondence}, bilingual German--French edition, 1972; French edition, 1979; Spanish, 1994; Italian, 1995).  
No English edition of this work exists, though an English-language account can be found in Christian Bracco’s article \emph{Einstein and Besso, from Zurich to Milano} \cite{Bracco2014}.  

Among Einstein’s intellectual companions, Michele Besso holds a singular place.  
An engineer by training, he combined technical competence, wide intellectual curiosity, and great personal discretion.  
For several years he was one of Einstein’s closest interlocutors in Bern, and the explicit acknowledgment at the end of the 1905 paper shows that his role was not merely social.  
Einstein regarded their discussions as having contributed directly to the clarification of the problem.

A later testimony gives this acknowledgment a more vivid form.  
According to the account of Einstein’s 1922 Kyoto lecture, Einstein said that after returning home from a discussion with Besso around mid-May 1905, he finally grasped the solution he had been seeking.  
The next day, he told Besso: ``I have completely solved the problem.'' Besso appears as a trusted interlocutor whose questions, objections, and technical understanding seem to have helped Einstein bring a long-standing problem to its final form.  

\subsection{Einstein’s Bibliographical Silence}

One of the most striking features of the 1905 paper is Einstein’s silence concerning much of the scientific work of the preceding two decades.  
Major experiments, such as those of Michelson--Morley, Trouton--Noble, Rayleigh, and Brace, are not mentioned.  
Nor are the recent theoretical works of Lorentz and Poincaré, both central figures in contemporary discussions of electrodynamics and the principle of relativity.  
A significant body of experimental and theoretical research, central to debates between the late 1880s and 1905, is absent from the explicit argumentative framework of the paper.
  
In 1905 Einstein did not cite Lorentz’s 1895 or 1904 memoirs and did not place Lorentz in the historical framing of the paper.  
Lorentz was mentioned only later, in the technical electrodynamical part, in connection with the Lorentzian theory of the electrodynamics and optics of moving bodies. The other few named references belonged mainly to older and already established traditions, such as Maxwell, Hertz, and Doppler.  
By contrast, Poincaré’s recent work, and the recent experimental literature on ether drift---Michelson--Morley, Trouton--Noble, Rayleigh, and Brace---were absent from the explicit argumentative framework of the paper.  
The result was a text that presented special relativity less as the outcome of an ongoing contemporary debate than as a conceptual clarification arising from general principles and classical electrodynamics.

This choice is all the more striking given the evidence that Einstein was, from 1900 onward, an active and attentive reader of recent scientific literature.  
His letters to Mileva Marić indicate that he used scientific libraries, mobilized personal networks to obtain books and articles, and explicitly sought recent works by Lorentz through his former classmate Jakob Ehrat.  
He also very plausibly encountered the 1900 \emph{Festschrift} in honor of Lorentz at the Istituto Lombardo, as Christian Bracco has argued.  
The portrait that emerges is not that of an isolated young researcher cut off from the contemporary literature, but of a curious and resourceful physicist able to use libraries, correspondence, and personal contacts to follow current developments.

This makes the silence of the 1905 paper historically significant. The contrast between Einstein’s documented reading practices and the sparseness of his references raises a legitimate historiographical question: how much of the contemporary electrodynamic context was silently absorbed into the conceptual architecture of the 1905 paper?

Einstein’s later statements make this question even more delicate.  
In several retrospective accounts, he suggested that the decisive step consisted in recognizing that Lorentz’s local time \(t'\) should be understood as the time of the moving system. Yet the 1905 paper itself gives no explicit account of such a transition from Lorentz’s work to Einstein’s own interpretation.  
The authors and phenomena named in the text do not provide the reader with the immediate source of the variable \(t'\), nor with the recent debates through which local time, length contraction, and the relativity principle had been discussed. There is therefore a tension between the later retrospective narrative, which places Lorentz’s local time at the center of the transition, and the 1905 paper, which presents the theory without recounting that recent intellectual background.

The title chosen by Einstein, \emph{Zur Elektrodynamik bewegter Körper} --- ``On the Electrodynamics of Moving Bodies'' --- adds another layer to this problem.  
Such phrasing was not new: the electrodynamics of moving bodies had already become a recognized field of research, associated with Maxwell, Hertz, Lorentz, Abraham, Cohn, Wien, Poincaré, and others \cite{Ginoux2024}.  
Einstein’s title therefore placed the paper within an existing tradition.  
Yet the paper itself opens not with a survey of that literature, but with the asymmetry between magnet and conductor, followed by a kinematic reconstruction of space and time.  
Electrodynamics proper appears mainly after this kinematic foundation has been established.  
The title thus connects the paper to a contemporary field, while the exposition largely detaches the argument from the detailed literature of that field.

The 1905 paper may therefore be understood as performing a powerful act of conceptual reorganization.  
It does not simply add another contribution to the existing electrodynamics of moving bodies; it reformulates the problem from the standpoint of principles, synchronization procedures, and inertial frames.  
This is part of its strength. At the same time, the absence of explicit references to the recent experimental and theoretical context contributed to the later impression that special relativity emerged almost independently of the dense literature that immediately preceded it.

As the French historian of science Michel Paty observed \cite{paty1993}, the genesis of special relativity retains an element of enigma:
\begin{quote}
``Einstein’s work seems to have appeared fully formed: no prior publications, and no preparatory manuscripts among his papers, except for one youthful essay.  
The genesis of special relativity seems shrouded in mystery.''
\end{quote}
  
The absence of preparatory traces, combined with the paper’s sparse bibliographical apparatus, is precisely what makes the question of Einstein’s sources, readings, and silences historically significant.

\subsection{The Poincaré--Einstein Chronology in June 1905}

The chronology of June 1905 deserves attention. On 5 June 1905, Henri Poincaré presented to the Paris Academy of Sciences his note \emph{Sur la dynamique de l'électron}, published in the \emph{Comptes rendus de l'Académie des sciences}.  
Einstein's paper on special relativity reached the \emph{Annalen der Physik} on 30 June 1905.  
The interval between these two dates is short, but historically significant.

It must first be emphasized that Einstein's work was already advanced before Poincaré's note appeared.  
In his letter to Conrad Habicht dated 18 May 1905, Einstein announced several papers then in preparation: one on radiation and light quanta, another on Brownian motion, and a third dealing with the electrodynamics of moving bodies and involving a modification of the theory of space and time. Einstein was already working on the problem in mid-May, and probably had a substantial draft or at least a clearly structured project by then.

The question raised here is therefore narrower.  
It concerns not the origin of Einstein's theory, but the possible timing of its submission.  
The issue is whether the appearance of Poincaré's note in the \emph{Comptes rendus}, if it became visible in Bern during June, may have encouraged Einstein to send an already advanced manuscript without further delay.

The material possibility of rapid circulation should not be dismissed.  
The \emph{Comptes rendus} were distributed quickly to scientific institutions, and the University Library of Bern regularly received the journal, as confirmed by surviving collections.  
It is therefore plausible that the issue containing Poincaré's note could have been available in Bern in the second half of June.  
This does not prove that Einstein saw it, but it removes any argument based on material impossibility.

A striking example of the speed of early twentieth-century scientific communication is provided by Einstein's own paper on the photoelectric effect.  
The manuscript bears the handwritten date ``Bern, 17 March 1905.''  
The printed journal states that it was received by the \emph{Annalen der Physik} on 18 March 1905---only one day later, at the Leipzig editorial office.  
For the relativity paper, by contrast, the printed journal gives only the date of receipt: 30 June 1905.

Einstein thus had, in principle, a short window during which Poincaré's note could have become known to him before the dispatch of his own manuscript. 
There is no direct evidence that Einstein read Poincaré's note before sending his paper.  

The important point is chronological. In the first half of 1905, Einstein was working at an extraordinary pace.  
He had already submitted the light-quantum paper in March and the Brownian-motion paper in May, prepared a new doctoral dissertation, and contributed bibliographical notices to the \emph{Beiblätter zu den Annalen der Physik}.  
This activity took place while he was employed full time at the Patent Office and while he had family obligations.  
The submission of the relativity paper at the end of June therefore occurred in a period of exceptional scientific intensity.

In this context, the publication of Poincaré's note may have had a limited but historically meaningful effect. If Einstein became aware of it in June, it could have acted as a signal that the electrodynamics of moving bodies had entered a moment of rapid publication and possible priority.  
It may therefore have encouraged him to submit an already advanced manuscript without waiting longer.

The University Library of Bern also made such awareness materially possible. The library was only a short walk from the Patent Office, where Einstein worked during the day. This shows that the practical conditions for consulting recent scientific journals in Bern were favorable.

\section{Einstein’s Attitude toward Poincaré after 1905}

In his writings and lectures on relativity, Einstein frequently cited Lorentz and Maxwell, but he almost always left Poincaré outside the scientific narrative of special relativity.  
Lorentz was generally presented by Einstein as the continuer of Maxwell’s work, incorporating electron theory into electromagnetism.  
The same line recurs in his major lectures of the 1920s.  
At Leiden in 1920, Einstein praised Lorentz's theory of the ether as the most important advance in the theory of electricity since Maxwell, and stated that the kinematics of special relativity had been modelled on Maxwell--Lorentz electrodynamics.  
At Princeton in 1921, he again presented special relativity as arising from the tension between the principle of relativity and Maxwell--Lorentz electrodynamics, whose equations were not covariant under the Galilean transformation but had proved successful in optics and electrodynamics.  
In his 1923 Nobel lecture, he formulated the point even more explicitly, describing special relativity as an adaptation of physical principles to Maxwell--Lorentz electrodynamics. 
In these texts, Poincaré is not mentioned, while Einstein repeatedly emphasizes the pivotal role of electrodynamics in the genesis of special relativity.

There is one important exception in a different domain.  
In 1906 \cite{einstein1906}, in a paper on the inertia of energy, Einstein explicitly referred to Poincaré’s 1900 work, where Poincaré had already stated that electromagnetic energy carries a ``fictitious mass.''  
Einstein acknowledged that the essential idea was already present in Poincaré, even if he preferred to propose his own demonstration.  
This is the only clear case known to me in which Einstein gave direct scientific credit to Poincaré. But on the core of special relativity itself, the silence remained striking.

In a tribute to Lorentz on the occasion of his death in 1928, Einstein exalted Lorentz’s scientific and moral stature, but added a small critical remark: ``H. A. Lorentz even found the transformation that bears his name, without noticing its group properties.''  
This statement is surprising.  
On the one hand, it is accurate: Lorentz had not identified the group structure.  
But Einstein himself, in 1905, had not examined it either, apart from a passing allusion.  
Only Poincaré, in 1905 and 1906, emphasized the group property, with a very detailed study in his 1906 paper.  
Yet again, Einstein said nothing about this in his homage to Lorentz.

The historian Abraham Pais highlighted Einstein’s silence regarding Poincaré.  
In his biography of Einstein \cite{Pais1982}, he recounts that in the 1950s he gave Einstein a copy of Poincaré’s 1906 memoir.  
Einstein replied that he did not know this text and would read it.  
After Einstein’s death, Pais asked Helen Dukas---Einstein’s longtime secretary and custodian of his papers---whether Poincaré’s article was among his archives; it was not.
Pais concluded laconically: ``The mystery of the Einstein--Poincaré connection remains.''

There is another register in which Einstein cited Poincaré: philosophy.  
In his 1921 Berlin lecture, he discussed Poincaré’s conventionalism concerning geometry.  
He acknowledged the force of the argument that geometrical laws are conventions, but he criticized it by emphasizing that general relativity imposes an effective geometry dictated by physical phenomena themselves.  
Here, Poincaré is explicitly mentioned, but to be contested rather than praised.  
At the meeting of 6 April 1922 of the \textit{Société française de philosophie}, during Einstein’s visit to Paris, the session’s chair, the philosopher Xavier Léon, was the only participant, in the published discussion, to invoke Poincaré’s work on electrodynamics and relativity.  
After recalling that Poincaré was one of the Society’s founders and a major figure in electrodynamics, Léon quoted a passage from the 1901 \emph{Leçons sur l'électricité et l'optique} \cite{poincare1901}.  
There, Poincaré formulated the hypothesis that optical phenomena depend only on the relative motions of material bodies, adding that a well-constructed theory should allow the principle to be demonstrated in full rigor, and not only to first order as in Lorentz’s 1895 paper, and that among existing theories Lorentz’s came closest.  
In his reply, Einstein did not enter into the details of this remark.  
He confined himself to situating Poincaré on philosophical ground, opposing his conventionalism to Kantian philosophy, and affirming his own middle path based on experience.

To my knowledge, during this Paris visit---when several days of debates were held in the presence of eminent French scientists---Léon’s intervention was the only explicit allusion, in the published record, to Poincaré’s work related to relativity.

A late and revealing document nevertheless complicates this long silence.  
At the end of 1953, André Mercier, then associated with theoretical physics at the University of Bern and one of the organizers of the forthcoming jubilee, invited Einstein to take part in the celebration of the fiftieth anniversary of special relativity.  
Einstein declined the invitation in a letter to Mercier dated 9 November 1953, citing reasons of health.  
He nevertheless added that he hoped care would be taken to ensure that the merits of H. A. Lorentz and H. Poincaré would also be suitably acknowledged on that occasion.  

Max Born did later give a lecture at the Bern jubilee \cite{MercierKervaire1956}, in which he discussed the contributions of both Lorentz and Poincaré.  
Probably this emphasis resulted from a specific request by Mercier or the organizing committee.
 
\section{Einstein’s 1907 Review Article on Special Relativity}

In his 1907 review article \cite{einstein1907}, Einstein presented his 1905 theory for the first time in an extended didactic form.  
This text is historiographically important because it does not merely explain special relativity; it also constructs an early retrospective narrative of its origin.  
In contrast with the 1905 paper, where no recent literature on the electrodynamics of moving bodies was cited, the 1907 review explicitly placed Lorentz’s 1904 memoir at the center of the story.

Einstein stated that Lorentz had shown how the Maxwell--Lorentz equations could account for the negative result of the Michelson--Morley experiment.  
He also emphasized, however, that Lorentz’s transformation still appeared within an ether-based theory and that the local time retained a special status distinct from true time.  
The essential step, in Einstein’s presentation, was therefore not the formal introduction of the transformation itself, but its reinterpretation: the local time of the moving system had to be understood as the time measured by clocks in that system.
  
It would be misleading to say that Lorentz’s \(t'\) was merely a mathematical variable without physical meaning.  
In Lorentz’s 1904 theory, local time was already connected with the behavior of electromagnetic systems in motion and with the explanation of the negative ether-drift experiments.  
Lorentz had also shown that electromagnetic clocks, or idealized electromagnetic oscillators, would be retarded in motion.  

Einstein’s 1907 review cited four pre-1905 external references:
\begin{itemize}
  \item A. A. Michelson and E. W. Morley (1887);
  \item H. A. Lorentz (1895);
  \item H. A. Lorentz (1904);
  \item E. C. T. Trouton and H. R. Noble (1903).
\end{itemize}
Einstein also cited his own 1905 paper.  
Thus, in 1907, Lorentz’s 1904 memoir became an explicit and central reference in Einstein’s retrospective exposition of special relativity.  
By contrast, Poincaré’s 1905 note and his 1906 Palermo memoir were not mentioned.

This contrast with the 1905 paper is striking.  
In 1905 Einstein did not cite Lorentz’s 1904 memoir.  
In 1907, however, Lorentz 1904 was presented as one of the essential antecedents for understanding the formal structure of the theory.  
Einstein also indicated that this memoir had not been known to him when he wrote his 1905 article.  
The review article had been requested from Einstein by Johannes Stark.  
In a letter to Stark dated 25 September 1907, Einstein accepted the request but warned that his bibliographical knowledge was limited by his working conditions at the Patent Office.  
He wrote that he was not in a position to acquaint himself with everything published on the topic, because the library was closed during his free time.  
He added that, apart from his own works, he knew one paper by H. A. Lorentz from 1904, one by E. Cohn, one by Mosengeil, and two by Planck, and asked Stark to inform him of any additional relevant publications.

The narrative constructed in 1907 would become highly influential.  
It presented the genesis of special relativity through a line running from Michelson--Morley and Lorentz to Einstein’s reinterpretation of space and time.  
This was a pedagogically effective account, but it was also selective.  
It acknowledged Lorentz’s decisive role, but left out Poincaré’s formulation of the principle of relativity, his interpretation of local time, his determination of the scale factor \(l(v)\), and his demonstration of the group property of the Lorentz transformations.

\subsection{Einstein's 1910 Review Article on the Principle of Relativity}

In 1910, Einstein published a new review article on relativity in French translation in the \emph{Archives des sciences physiques et naturelles}, under the title \emph{Le principe de relativité et ses conséquences dans la physique moderne} \cite{einstein1910}.  
The article was not a mere summary of the 1905 paper, but a detailed pedagogical reconstruction of the emergence of special relativity for a French-speaking scientific readership.  
Einstein opened with the ether, discussed the optics of moving bodies, Fizeau's experiment, Lorentz's electron theory, the negative ether-drift experiments, Michelson--Morley, the Lorentz--FitzGerald contraction, the magnet-and-conductor asymmetry, the principle of relativity, the abandonment of the ether, and the operational definition of time by means of clocks.  
In the second part, he introduced the Lorentz transformations, interpreted length contraction and clock retardation as consequences of the theory, incorporated Minkowski's space--time formalism, and briefly turned to mechanics, citing Stark's 1906 work.

Lorentz occupied the central place among the pre-Einsteinian theoretical antecedents.  
Einstein explicitly cited Lorentz's 1895 theory, in its 1906 edition, and stated that the space--time transformations had been introduced ``in a very fortunate way'' into electrodynamics by Lorentz, before designating them as the ``Lorentz transformations.''  
He also presented the Lorentz--FitzGerald contraction, formerly introduced to account for the negative result of the Michelson--Morley experiment, as a natural consequence of the principles of relativity and light-speed constancy.  
Poincaré, by contrast, was not mentioned.

This omission is particularly striking in view of the context and content of the publication.  
Einstein's review appeared in French and reconstructed in detail the emergence of special relativity through precisely the themes on which Poincaré had repeatedly intervened between 1898 and 1906: the measurement of time, simultaneity, Lorentz's local time, the principle of relativity, the critique of the ether, the impossibility of detecting absolute motion, the Lorentz transformations, and the dynamics of the electron.  
Yet Poincaré remained entirely absent from the account.

The 1910 case is also significant because Einstein was no longer in the same institutional situation as in 1907.  
By then he had left the Patent Office and had entered academic life in Zurich.  
He had moved to Zurich in 1909, after being appointed extraordinary professor of theoretical physics at the University of Zurich, with regular access to university resources and to a fuller academic environment.  
The continued absence of Poincaré from the 1910 review therefore cannot be explained simply by the immediate practical constraints that Einstein had invoked in his 1907 letter to Stark.

Taken together, the 1907 and 1910 reviews mark an important stage in the formation of Einstein's public narrative of special relativity.  
Einstein explicitly recognized Lorentz as the central electrodynamical antecedent and presented special relativity as the conceptual resolution of the difficulties internal to the Lorentzian ether theory.  
But Poincaré remained outside the story.  
The result was a genealogy in which Lorentz supplied the mathematical and electrodynamical background, Einstein supplied the conceptual reinterpretation, Minkowski supplied the space--time formalism, and Poincaré's role disappeared from Einstein's own public account.

\subsection{Einstein’s 1922 Kyoto Lecture}

In his 1922 Kyoto lecture, according to the later account preserved from the event, Einstein presented a retrospective narrative of the genesis of special relativity.  
He recalled three main elements: his youthful thought experiment of pursuing a light wave, the asymmetry between the descriptions of a moving magnet and a moving conductor, and the decisive role of his discussions with Michele Besso.  
In this account, Einstein explained that the crucial step consisted in recognizing the physical meaning of time in a moving system.  
Lorentz’s local time was no longer to be regarded as a secondary or merely auxiliary quantity; it had to be understood as the time actually indicated by clocks in the moving system.

This testimony is important, but it must be used with caution.  
It is a retrospective narrative given many years after the events, and it compresses into a few dramatic moments a much longer intellectual process.  
It nevertheless became one of the most influential sources for the image of Einstein’s discovery as a sudden conceptual breakthrough: after years of difficulty, the solution appeared through the recognition of the relativity of simultaneity and the physical interpretation of time.

If one compares Einstein’s trajectory with those of Lorentz and Poincaré, a striking difference appears.  
Lorentz had been engaged since the 1890s in studying the electrodynamics of moving bodies.  
Starting in 1892, he developed his electron theory.  
In 1895, he introduced local time.  
He gradually refined his equations up to his 1904 paper, where the Lorentz transformations appeared almost in their final form.  
Each step was published, discussed, criticized, and corrected: one can follow a continuous sequence of partial results, the fruit of more than a decade of work.

Poincaré, for his part, began as early as 1898 to address the question of the measurement of time and simultaneity.  
In 1900, in \emph{La théorie de Lorentz et le principe de réaction}, he discussed the principle of reaction, electromagnetic momentum, and the difficulties raised by Lorentz’s electrodynamics.  
Between 1902 and 1904, his books and lectures asserted with increasing clarity the central role of the principle of relativity, and in 1904 he gave it a general statement.  
His 1905 note and 1906 memoir marked new thresholds: the group property of the Lorentz transformations, the determination of Lorentz’s scale factor, the transformation of charge and current densities, the law of velocity addition, and the use of the principle of least action within a Lorentz-invariant framework.  
Here again, one observes a documented progression, marked by successive publications.

The contrast with Einstein is striking.  
Before 1905, there are signs of his interest in electrodynamics, the ether, Michelson’s experiment, relative motion, and Lorentz’s work.  
His letters show that these questions had occupied him for several years.  
But there is no surviving preparatory manuscript of the 1905 relativity paper, no series of partial publications on the problem, and no preserved chain of calculations comparable to the visible trajectories of Lorentz and Poincaré.  
By mid-May 1905, as his letter to Conrad Habicht shows, Einstein was already preparing a paper on the electrodynamics of moving bodies involving a modification of the theory of space and time.  
A few weeks later, the completed article was submitted to the \emph{Annalen der Physik}.  
The archival record therefore moves abruptly from scattered earlier indications to the finished paper.

It is precisely this discontinuity in the surviving record that nourished the image of a lightning-like act of genius.  
John Stachel \cite{Stachel1986} emphasized the extraordinary rapidity with which Einstein, after the decisive step, drew the consequences and sent his paper to the \emph{Annalen der Physik}.  
John Norton \cite{Norton2014} likewise analyzed the decisive role of the relativity of simultaneity and the compressed period of writing that followed. Gerald Holton \cite{holton1973} also emphasized the distinctive simplicity of Einstein’s 1905 move: starting from two principles, elevating them to foundations, and deriving their consequences.

These commentaries do not all claim that Einstein worked in isolation, nor that special relativity emerged without antecedents.  
But the structure of the surviving evidence makes such an image easy to construct.  
Lorentz and Poincaré left visible sequences of publications leading step by step toward relativity.  
Einstein, by contrast, left almost no preparatory trace for the decisive paper.  
The result is a historiographical asymmetry: the documented gradual construction of Lorentz and Poincaré stands beside the apparently sudden completion of Einstein’s 1905 article.  
This asymmetry has contributed powerfully to the enduring narrative of a solitary and almost instantaneous breakthrough.

\section{The 1913 Note and the Canonization of the Lorentz--Einstein Line}

In 1913, the German publisher B. G. Teubner in Leipzig and Berlin issued a volume entitled \emph{Das Relativit\"atsprinzip: Eine Sammlung von Abhandlungen}.  
The book appeared with a foreword by Otto Blumenthal and notes by Arnold Sommerfeld.  
It brought together a small set of texts that were already being treated as foundational for the new theory of relativity.  
The names printed at the head of the volume were H. A. Lorentz, A. Einstein, and H. Minkowski.  
This editorial framing is already significant: it presented relativity through the sequence Lorentz--Einstein--Minkowski, while Henri Poincaré was absent as an author of the collection.

The first edition contained the following texts:
\begin{itemize}
  \item H. A. Lorentz, ``Der Interferenzversuch Michelsons'' (1895);
  \item H. A. Lorentz, ``Elektromagnetische Erscheinungen in einem System, das sich mit beliebiger, die des Lichtes nicht erreichender Geschwindigkeit bewegt'' (1904);
  \item A. Einstein, ``Zur Elektrodynamik bewegter K\"orper'' (1905);
  \item A. Einstein, ``Ist die Tr\"agheit eines K\"orpers von seinem Energieinhalt abh\"angig?'' (1905);
  \item H. Minkowski, ``Raum und Zeit'' (1909), based on the Cologne lecture delivered in 1908, followed by Sommerfeld's notes (1913);
  \item H. A. Lorentz, ``Das Relativit\"atsprinzip und seine Anwendung auf einige besondere physikalische Erscheinungen'' (1910).
\end{itemize}

The volume thus offered a compact genealogy of special relativity.  
It was not a neutral anthology of all relevant contributions, but a selective canon.  
The path proposed to the reader led from Lorentz's analysis of Michelson's experiment and his 1904 electron theory to Einstein's reformulation, then to Minkowski's space--time interpretation, before returning to Lorentz's later exposition of the relativity principle.  
Poincaré's 1905 note and 1906 Palermo memoir were not included as primary texts.  
This absence is especially striking in relation to Minkowski's 1908 lecture \emph{Raum und Zeit}.  
Minkowski did not mention Poincaré there, despite the formal proximity between some of Poincaré's results and the mathematical structures that Minkowski would geometrize.  
Poincaré had not created Minkowski's geometrical space--time in the full sense, but he had already emphasized the Lorentz group, the invariant quadratic form, and a four-dimensional formal analogy in the formulation of relativistic electrodynamics.  
Sommerfeld's notes did acknowledge some of these anticipations, especially in relation to Poincaré's 1906 Palermo memoir.  
But this acknowledgement remained confined to the apparatus of commentary.  
Poincaré did not become one of the authors through whom the collection organized the origins of relativity.  
For this reason, the editorial structure of the volume is revealing: it could absorb some mathematical elements associated with Poincaré while leaving him outside the main canonical line.

A particularly revealing element appears at the beginning of Einstein's 1905 paper as reprinted in the 1913 volume.  
Attached to the introductory reference to results already established to first order, the following note was added:
\begin{quote}
``Die im Vorhergehenden abgedruckte Arbeit von H. A. Lorentz war dem Verfasser noch nicht bekannt.''
\end{quote}
In translation:
\begin{quote}
``The work by H. A. Lorentz printed immediately above was not yet known to the author.''
\end{quote}

This note is highly unusual.  
It was added eight years after the publication of Einstein's 1905 article.  
The note says that the Lorentz paper printed immediately before Einstein's article in the same volume was not yet known to Einstein when he wrote his own work.  
In the Teubner volume, the reader encounters Lorentz's 1904 memoir before Einstein's 1905 paper.  
This editorial order naturally suggests a historical sequence: Lorentz first, Einstein second.  
The note therefore interrupts a possible inference produced by the anthology itself, namely that Einstein had written his paper after reading Lorentz's detailed 1904 memoir.

The context helps explain why such a clarification may have seemed necessary.  
From 1907 onward, the expression ``Lorentz--Einstein theory'' began to circulate in German scientific circles.  
A private early attestation occurs in a letter from Max Planck to Wilhelm Wien dated 24 May 1907, in which Planck wrote that he was working on the Lorentz--Einstein theory of relativity.  
A few weeks later, Planck presented his results before the Prussian Academy, on 13 June 1907.  
The terminology indicates that, at least in some German contexts, the new theory was not yet perceived as Einstein's theory alone, but as a framework associated with both Lorentz and Einstein.

The expression soon entered the printed literature.  
In 1908, Alfred Bucherer published in the \emph{Physikalische Zeitschrift} an article entitled ``Measurements on Becquerel Rays: Experimental Confirmation of the Lorentz--Einstein Theory.''  
This was one of the earliest explicit public uses of the expression as a scientific label.  
In the following years, similar terminology spread further.  
In 1910, Tullio Levi-Civita published in the \emph{Annalen der Physik} an article on the rigid motions of Lorentz--Einstein, where the adjective referred to an already established kinematic framework.  
Emil Cohn also used the expression in his lectures on the principle of relativity, later published in 1913.

These examples show that, by the late 1900s and early 1910s, the expression ``Lorentz--Einstein'' circulated across private correspondence, major physics journals, and broader expository works.  
It suggests that relativity was still often perceived as having a shared intellectual parentage, at least between Lorentz and Einstein.  
By contrast, Poincaré, although deeply involved in the same theoretical developments, was largely excluded from this emerging editorial and terminological tradition.

The German collection did not remain fixed in its original form.  
Later editions enlarged it in the direction of general relativity.  
The fourth German edition of 1922, translated into English by Methuen in 1923 under the title \emph{The Principle of Relativity}, no longer simply reproduced the 1913 arrangement.  
Lorentz's 1910 Göttingen lecture was no longer included, while several later papers by Einstein on gravitation and general relativity, together with Hermann Weyl's paper on gravitation and electricity, were added.  
Yet the opening sequence devoted to the origins of special relativity remained unchanged: Lorentz's analysis of Michelson's experiment, Lorentz's 1904 memoir, Einstein's two papers of 1905, and Minkowski's \emph{Raum und Zeit}.  
In this sense, the basic editorial pattern remained intact: the genealogy of special relativity was still presented through the Lorentz--Einstein--Minkowski line, with Poincaré absent as an author of the collection.

The 1923 Methuen edition reinforced this pattern for an Anglophone readership.  
Sommerfeld's notes did mention Poincaré, especially in relation to four-dimensional formalism and relativistic dynamics, but these acknowledgements remained confined to the apparatus of commentary.  
No work by Poincaré was included, and he did not become one of the principal figures through whom the collection organized the origins of special relativity.  
The Dover edition of 1952, still widely circulated today, reproduced the 1923 translation and thus helped preserve the same canon \cite{einstein-lorentz2000}.

The note added to Einstein's 1905 paper in the German collection of 1913 also became part of this editorial tradition.  
It was retained in later German editions and was carried over into the English Methuen edition of 1923, where it appeared as:
\begin{quote}
``The preceding memoir by Lorentz was not at this time known to the author.''
\end{quote}
The note therefore did not remain an isolated clarification, but acquired a stable place in the documentary canon through which many readers encountered the early history of relativity.

The 1913 note thus fits into a broader historiographical situation.  
It acknowledged, indirectly, that the relation between Lorentz's 1904 memoir and Einstein's 1905 paper required clarification.  
But it did so inside an editorial tradition in which the relevant prehistory of special relativity had already been narrowed to Lorentz, Einstein, and Minkowski.  
The note is therefore doubly revealing: it shows that the Lorentz--Einstein relation already needed careful framing in 1913, and it leaves untouched the more complete question of Poincaré's exclusion from the canonical narrative.

\section{The Absence of Any Review of Poincaré’s 1905 and 1906 Papers in the \emph{Beiblätter zu den Annalen der Physik}}

It is striking that Poincaré’s two main technical papers on relativity, the 1905 note \emph{Sur la dynamique de l’électron} and the 1906 Palermo memoir of the same title, appear to have received no review in the \emph{Beiblätter zu den Annalen der Physik}.  
Christian Marchal \cite{marchal2004} had already pointed out this omission, noting at the same time that other works by Poincaré from the same period were indeed summarized in that journal.  
This point can now be checked more directly.  
The relevant volumes of the \emph{Beiblätter}, together with their cumulative indexes for 1892--1906 and 1907--1919, are available in digitized form.  
A combined examination of the indexes and full-text searches has revealed no notice or abstract devoted to either Poincaré’s 1905 note or his 1906 Palermo memoir.

The omission is significant.  
The \emph{Beiblätter} reviewed foreign scientific literature extensively, including works published in French and Italian journals.  
They also reviewed other publications by Poincaré.  
The absence of his two relativity papers therefore cannot be explained simply by linguistic barriers or by a general neglect of Poincaré as an author.  
It points instead to a more specific failure of bibliographical transmission: two texts that are now central to the history of relativity did not enter one of the principal German review channels through which physicists followed recent international literature.

This absence had consequences.  
A German-speaking reader who relied on the \emph{Beiblätter} in 1905--1906 could encounter Gans’s clear summary of Lorentz’s 1904 memoir, and could follow parts of the German debate on the electrodynamics of moving bodies.  
But he would not have been alerted, through this same channel, to Poincaré’s 1905 note or to the much fuller 1906 memoir in the \emph{Rendiconti del Circolo matematico di Palermo}.  
The \emph{Beiblätter} therefore contributed, unintentionally but materially, to an asymmetry of visibility: Lorentz’s 1904 results were summarized in German, while Poincaré’s corresponding contributions were not.

\section{Wilhelm Wien Nominates Albert Einstein for the 1912 Nobel Prize in Physics}

In 1912, Wilhelm Wien, German physicist and recipient of the 1911 Nobel Prize in Physics, submitted a nomination to the Nobel Committee that included Albert Einstein \cite{Nobel1912Wien}.  
The official Nobel nomination record shows, however, that this was not a nomination of Einstein alone.  
Wien's first choice was P. N. Lebedev for an undivided prize; as a further proposal, he nominated Albert Einstein and H. A. Lorentz for a divided prize.

Wien's justification was remarkably explicit.  
From a purely logical point of view, he described relativity theory as one of the most important achievements of theoretical physics.

Wien's nomination placed Einstein and Lorentz together in the context of the relativity principle, reflecting the contemporary perception of the theory as still strongly connected with Lorentz's electrodynamics.  
It confirms that, in the early 1910s, Einstein's contribution to relativity could still be understood in relation to Lorentz's mathematical and physical framework rather than as an entirely isolated creation.

Poincaré was not included in Wien's nomination.  
This absence is especially notable because Poincaré did receive other nominations for the 1912 Nobel Prize in Physics.  
In the specific German context represented by Wien's nomination, however, the relevant pairing was Einstein and Lorentz.  
This is consistent with the broader pattern examined in this article: in German-speaking scientific and institutional contexts, the emerging public memory of relativity was increasingly organized around the Lorentz--Einstein line, while Poincaré's role remained marginal.

\section{Wolfgang Pauli’s Remarkable Review Article}

In 1921, at only twenty-one years of age, Wolfgang Pauli published a comprehensive review article on relativity in the German \emph{Encyklopädie der mathematischen Wissenschaften}.  
The article, entitled \emph{Relativitätstheorie}, had been entrusted to him by Arnold Sommerfeld, who was then heavily occupied with other problems and recognized the exceptional abilities of his young student.  
Although Pauli was still at the beginning of his academic career, the result quickly became one of the most authoritative expositions of both special and general relativity.  

\subsection*{Scientific Content}

Pauli presented with exceptional rigor the main structures of the theory:
\begin{itemize}
  \item the kinematics of special relativity, including Lorentz transformations, simultaneity, and velocity composition;
  \item the formal unification of relativistic quantities through four-vectors and tensors;
  \item relativistic electrodynamics and the mechanics of the electron;
  \item the mathematical foundations and physical consequences of general relativity.
\end{itemize}

\subsection*{Poincaré’s Place}

One of the most remarkable features of Pauli’s article is the substantial place it gives to Poincaré.  
Poincaré is mentioned eleven times, and these references are not incidental.  
They concern some of the central mathematical and conceptual structures of special relativity. According to Yves Gingras \cite{Gingras2008}, Felix Klein made sure that Poincaré was adequately cited in Pauli’s article for the \emph{Encyklopädie der mathematischen Wissenschaften}.

Pauli explicitly recognized that Poincaré had filled important formal gaps left by Lorentz’s work.  
He credited him with having stated the principle of relativity in a general and rigorous form, and with having required that the laws of nature be covariant with respect to what Poincaré called the Lorentz transformation.  
Pauli also emphasized that Poincaré derived the invariance of the transverse dimensions from the group property of the transformations, corrected Lorentz’s formulae for charge density and current, and obtained the complete covariance of the field equations of electron theory.

This recognition is especially significant because Pauli also noted that the terms ``Lorentz transformation'' and ``Lorentz group'' first occurred in Poincaré’s work.  
Thus, Poincaré appears in Pauli’s exposition as an important contributor to the mathematical formulation of special relativity.

At the same time, Pauli did not erase the difference between Poincaré and Einstein.  
He clearly distinguished the approach of Lorentz and Poincaré, who took Maxwell’s equations as the basis of their considerations, from Einstein’s more radical foundation.  
For Pauli, Einstein’s distinctive achievement was to derive the covariance law from simpler basic assumptions, especially the constancy of the velocity of light and the relativity principle.  
The point is therefore not that Pauli assimilated Poincaré to Einstein, but that he gave Poincaré a real and substantial place in the genealogy of the theory.

\subsection*{Historiographical Significance}

The historiographical significance is considerable.  
In contrast with the Teubner--Methuen canon, organized around Lorentz, Einstein, and Minkowski, Pauli’s article reintegrated Poincaré into the mathematical genealogy of special relativity.  
This reintegration concerned precisely the points on which Poincaré’s contribution had been most important: the principle of relativity, the group structure of the Lorentz transformations, the invariant formulation of electrodynamics, and the correction of Lorentz’s transformation laws for charge and current.

Pauli’s article therefore shows that Poincaré’s exclusion from the earlier Lorentz--Einstein--Minkowski canon was not inevitable.  
Within an authoritative German-language encyclopedic work, Poincaré could be treated as an essential figure in the mathematical construction of relativity.  
The contrast with the Teubner and Methuen collections is therefore striking: in those volumes Poincaré is absent, whereas in Pauli’s account he reappears at several structurally important points.

The article also stands out for the extraordinary maturity of its author.  
At only twenty-one, Pauli displayed exceptional technical and conceptual mastery.  
He did not merely summarize existing literature; he organized it critically, with a breadth and precision that made the article a lasting reference.  
Its formal clarity, analytical depth, and careful historical attributions explain why it shaped the teaching and presentation of relativity for decades.  
An English translation of this work appeared in 1958 \cite{pauli1958}, the year of Pauli’s death in Zurich.

\section{Richard Feynman and the Principle of Relativity: Another Notable Exception}

In many major physics textbooks of the twentieth century, the principle of relativity is presented primarily as one of Einstein’s two postulates, with no discussion of Henri Poincaré’s earlier formulation.  
Poincaré generally appears much more prominently in specialized works on the history of science than in standard pedagogical accounts of relativity.

A notable exception is found in Volume~I of Richard Feynman’s celebrated \emph{Lectures on Physics}.  
In the chapter on relativistic energy and momentum, Feynman explicitly recalls Poincaré’s formulation of the principle of relativity: the laws of physical phenomena must be the same for a fixed observer and for an observer in uniform translational motion, so that no experiment can reveal whether one is carried along in such a motion \cite{feynman-lectures}.

This acknowledgment is significant.  
Feynman does not merely mention Poincaré in passing; he reproduces Poincaré’s statement of the principle itself.  
Within the context of major physics teaching texts, this is a notable exception to the usual Einstein-centered presentation of special relativity.  
It shows that Poincaré’s priority in formulating the principle of relativity could be recognized even in a broadly pedagogical account, although such recognition remained uncommon.

\section{Some Final Considerations}

Poincaré and Einstein share the same fundamental starting point: the principle of relativity, understood as the equivalence of all inertial reference frames for the formulation of physical laws.  
Poincaré stated this principle explicitly in his 1905 note to the \emph{Comptes rendus}, where he treated it as a general postulate governing the laws of physics.  
Einstein made the same principle the first postulate of his 1905 paper.

The second element, however, is introduced in different ways.  
For Poincaré, the point of departure is the Maxwell--Lorentz theory and the requirement that the equations of electrodynamics satisfy the postulate of relativity.  
From this requirement, together with the group property of the transformations, one obtains the Lorentz transformations with \(l=1\), the relativistic law of velocity addition, and the invariant role of the velocity \(c\).  
In this sense, the invariance of \(c\) is embedded in the covariance structure of Maxwell--Lorentz electrodynamics.

Einstein reversed the order of presentation.  
He postulated directly that light in vacuum propagates with the same velocity \(c\) in every inertial frame, independently of the motion of the source.  
Combined with the relativity principle, this postulate leads to the Lorentz transformations and to the relativistic addition law for velocities.  
The covariance of the Maxwell--Lorentz equations then appears as a consequence of the new kinematics.  
Several commentators have noted this inversion: what is derived from electrodynamics in the Lorentz--Poincaré route becomes, in Einstein’s exposition, part of the kinematic foundation.

The result is a deep formal correspondence.  
In both cases, the same mathematical structures emerge: the Lorentz transformations, the velocity addition law, the invariant role of \(c\), and the covariance of Maxwell’s equations.  
The difference is not in the formal structure itself, but in the order of justification and in the conceptual interpretation attached to it.  
Poincaré’s construction remains embedded in a Lorentzian framework with an undetectable ether and a distinction between true time and local time.  
Einstein’s construction removes the privileged ether frame from the foundations and identifies the time of each inertial frame with the time measured by clocks in that frame.

For this reason, Einstein’s approach can rightly be called kinematic, but not in the sense of being historically detached from electrodynamics.  
The invariant velocity that he places at the foundation of the theory is not introduced abstractly as an unspecified limiting speed.  
It is the velocity of light in vacuum, known through Maxwell’s electrodynamics.  
Thus Einstein’s kinematics is universal in its ambition, but it emerges from the crisis of Maxwellian electrodynamics and from the impossibility of reconciling that theory with Galilean kinematics.

This point is important. A fully abstract kinematic formulation would have begun by postulating that there exists an invariant limiting velocity in nature.  
If this velocity were infinite, one would recover Galilean kinematics; if finite, one would obtain relativistic kinematics.  
Experiment would then identify this limiting velocity with the speed of light.  
Einstein did not proceed in this later axiomatic way.  
In 1905, he fixed the invariant velocity directly as the velocity of light, thereby rooting his kinematic reconstruction in electrodynamics itself.

Poincaré’s framework is less economical and less transparent than Einstein’s, because it retains the ether and the distinction between true time and local time.  
But this does not make it incoherent.  
Within Poincaré’s theory, the ether is dynamically undetectable, local time has an operational meaning through light-signal synchronization, and the observable laws obey Lorentz covariance.  
For the electromagnetic and kinematic phenomena under discussion, the Lorentz--Poincaré formulation and Einstein’s formulation lead to the same observable consequences, even though their conceptual interpretations differ.

Einstein’s own later expositions confirm that his theory remained closely tied to electrodynamics.  
In his 1907 review article, he emphasized that optical experiments show the propagation speed of light in vacuum to be constant and independent of the motion of the source.  
But he also connected this property with Maxwell--Lorentz electrodynamics and the principle of relativity.  
If Maxwell’s equations retain their form in all inertial frames, and if they imply waves propagating with velocity \(c\), then the same value of \(c\) must appear for all inertial observers.  
The constancy of light is therefore not an isolated intuition detached from electrodynamics; it is the point at which Maxwell’s theory and the relativity principle come into direct contact.

In summary, Poincaré and Einstein reached the same formal structure through different logical routes.  
For Poincaré, relativity was grounded in the covariance of Maxwell--Lorentz electrodynamics and in the requirement that the principle of relativity hold universally.  
For Einstein, relativity was presented as a kinematic reconstruction based on two postulates, but the second postulate took its content from Maxwell’s theory of light.  
The two formulations are therefore not identical in interpretation, but they are closely equivalent at the level of their mathematical structure and observable consequences.

This difference in presentation had enormous historical consequences.  
Einstein appeared as the founder of a new theory by principles: simple, economical, and conceptually transparent.  
Poincaré, by contrast, was often classified as a theorist of the Lorentzian electron tradition, burdened with the residual language of ether and true time.  
Yet this contrast can be misleading if it obscures the depth of Poincaré’s contribution: the explicit postulate of relativity, the group property of the Lorentz transformations, the determination \(l=1\), the invariant formulation of electrodynamics, and the use of the principle of least action within a Lorentz-invariant framework.

One may therefore argue that the deeper revolution from which special relativity emerged was the nineteenth-century electrodynamics of Ampère, Faraday, and Maxwell.  
It was Maxwell’s theory that introduced the field, fixed a universal electromagnetic velocity, and forced the conflict with Galilean kinematics.  
Special relativity did not arise \emph{ex nihilo}; it was the conceptual crystallization of a crisis already present in electrodynamics.  
Einstein’s achievement was to give this crisis its simplest and most powerful formulation.  
Poincaré’s achievement was to bring the Lorentzian programme to a high level of mathematical generality and to recognize the principle of relativity as a universal constraint on physical laws.

Thus, while Einstein’s formulation has the elegance of a principle-based theory and Poincaré’s retains a more elaborate conceptual apparatus, both belong to the same electrodynamic transformation of physics.  
The true historical picture is not one of an isolated creation opposed to a failed precursor, but of two closely related formulations emerging from the same Maxwellian revolution.

\end{document}